\newcommand{\PRL}[3]{Phys.\ Rev.\ Lett.\ {\bf #1},\ #2 (#3)}

\newcommand{\SC}[3]{Science\ {\bf #1},\ #2 (#3)}

\newcommand{\PRA}[3]{Phys.\ Rev.\ A\ {\bf #1},\ #2 (#3)}

\newcommand{\appsection}[1]{\let\oldthesection\thesection
  \renewcommand{\thesection}{ \oldthesection}
  \section{#1}\let\thesection\oldthesection}
\documentclass[twocolumn,preprintnumbers,amsmath,amsfonts,amssymb,notitlepage,showpacs,pra]{revtex4-1}
\usepackage{geometry}
\date{\today}
\DeclareMathAlphabet{\mathpzc}{OT1}{pzc}{m}{it}
\usepackage[utf8x]{inputenc}
\usepackage{amsmath}
\usepackage{amsfonts}
\usepackage{amssymb}
\usepackage{graphicx}
\usepackage{nicefrac}
\usepackage{amsbsy}
\usepackage{float}
\usepackage{epstopdf}
\usepackage{dsfont}
\usepackage{siunitx}

\def \be{\begin{equation}}
\def \ee{\end{equation}}
\def \ba{\begin{array}}
\def \ea{\end{array}}
\def \bea{\begin{eqnarray}}
\def \eea{\end{eqnarray}}
\def \myperp{\! \perp}

\begin{document}
\title{Stability of a Bose-Einstein condensate in a driven optical lattice: Crossover between weak and tight transverse confinement}
\author{Sayan Choudhury}
\email{sc2385@cornell.edu}
\author{Erich J Mueller}
\email{em256@cornell.edu}
\affiliation{Laboratory of Atomic and Solid State Physics, Cornell University, Ithaca, New York}
\begin{abstract}
We explore the effect of transverse confinement on the stability of a Bose-Einstein condensate (BEC) loaded in a shaken one-dimensional or two-dimensional square lattice. We calculate the decay rate from two-particle collisions. We predict that if the transverse confinement exceeds a critical value, then, for appropriate shaking frequencies, the condensate is stable against scattering into transverse directions. We explore the confinement dependence of the loss rate, explaining the rich structure in terms of resonances.
\end{abstract}

\maketitle
\section{Introduction}
Intense effort has been directed at using periodic driving to change the properties of cold atom systems \cite{HolthausFloquetReview,polkovnikov2014floquet}. For example, several experimental groups have been able to control effective hopping matrix elements by lattice shaking \cite{FerrariACTunnelingNatPhys2009,TinoNJP2010, TinoPRL2008,SengstockAbelianGaugePRL2012,Blochbandtopology, Ketterlespinorbit, BlochHarper,SengstockFrustratedScience2011,SengstockIsing2013}. The Sengstock group explored the physics of classical frustrated magnets \cite{SengstockFrustratedScience2011,SengstockIsing2013}. The Ketterle group saw evidence for Hofstader butterfly physics \cite{KetterleHarper,KetterleHarper2015}. The Esslinger group realized the topological Haldane model \cite{EsslingerHaldaneFloquet}. The list of experiments and theoretical proposals is large \cite{ChinFloquet2013,ChinFloquet2014,SengstockFloquet2015,EngelsFloquet2015,SengstockBerryCurvature,SengstockNonAbelianGaugePRL2012, ZhaiFTIarxiv, DemlerMajoranaPRL2011,DemlerFloquetTransport, OhMajoranPRB2013, BaurCooperFloquet2014,EckardtACEPL,DemlerFloquetAnamolous, ZhaoPRLAnamolous,MuellerFloquetAnamolous,creffieldsmf,GoldmanDalibardprx,PolkovnikovAnnals2013Floquet,chinz2theoryprl,
EckardtFloquetPRL2005,Gomez-LeonFloquetDimensionPRL,TorresPRBGraphene2012,TorresPRBGraphene2014,
GalitskiFTINatPhys2011,GaliskiFTIPRB2013,PodolskyFTI2013,BarangerFloquetMajorana,KunduSeradjehMajoranaPRL,
Cooperdalibardarxiv2014,Demlerarxiv2015,GalitskiFloquet2015,EckardtFloquetHighFrequency,floquetunitary,
DasRoyPRB2015,NeupertFloquetPRL2014,RefaelFAIarxiv2015,RigolChernNoGo2014,PontePRLMBL,DoraFloquetPRL2015}. There are, however both conceptual and practical issues with using periodic driving to control a system. A driven system has no ``ground state" nor a well-defined thermodynamic temperature. Furthermore, nearly all successful examples of this technique study non-interacting or very weakly interacting particles, and one almost always sees strong heating effects when moderate or strong interactions are introduced. Here we model some simple examples where these fundamental and practical issues are transparent. We make a series of predictions which are readily verifiable using techniques demonstrated in recent experiments \cite{ChinFloquet2013,ChinFloquet2014}, and which will enable the experimental study of interacting Floquet systems with cold atoms. \\

In earlier works we began studying these questions by modeling two geometries. First, we considered a 1D gas of atoms trapped in a shaken 1D lattice \cite{ChoudhuryMuellerPRA2014}. There we found large parameter regions where a Bose-Einstein condensate (BEC) is stable against 2-body collisions. Second, we considered a 3D gas of atoms trapped in a shaken 1D lattice, making an array of ``pancakes" \cite{ChoudhuryMuellerPRA2015}. We found that two-body collisions allowed energy to be taken from the shaking and transferred to transverse motion. The heating rates were consistent with those observed in experiment. In this setting, there is no steady state: the energy increases monotonically with time. The natural question is how these limits are connected. A 3D gas with harmonic transverse confinement should interpolate between these behaviors. Here, we calculate heating rates in this crossover. \\

We find a rich structure. First, there is a critical strength of the transverse confinement beyond which two-body collisions are unable to deplete the condensate. Second, as a function of the transverse confinement, the dimensionless loss rate is non-monotonic, displaying drops and jumps characteristic of resonances. We explain this behavior in terms of the opening and closing of transverse decay channels. Our results will be crucial to the next generation of experiments. For example, one will be unable to observe a Floquet fractional quantum Hall effect without tuning to parameters where losses are negligible. While other authors have conducted related studies of the stability of driven systems \cite{AbaninHeating2015,RigolFloqetPRX2014Floquet,EckardtPRLBoseSelection2013,ReafelFloquetBath,drivendissipativenatcomm2015,
RoschFloquetBoltzmann2015,ChoudhuryMuellerPRA2015,ChoudhuryMuellerPRA2014,Cooperarxiv2014,Cooperarxiv2015,CreffieldPRA2008,DemlerBukovPRL}, the question of collisional loss into transverse channels is relatively unexplored. While we focus on a particular model, the loss into transverse modes is quite generic in cold atoms (see for example \cite{ChoudhuryMuellerPRA2015}).\\

Driven systems are also of interest outside of cold atoms. For example, Photonics experiments have seen analogs of topological insulator physics in geometries which mimic electrons in a honeycomb lattice subjected to appropriate driving \cite{RechtsmanPFTI}. There are no interactions here. Translating these ideas to an electronic setting will require understanding the sort of loss processes which we study. Given the different structure of the interactions, our results cannot be applied directly in the solid state setting, but our logical framework is valuable.\\

Figure 1(a) depicts a 1-D lattice with weak transverse confinement yielding an array of pancake traps. We consider driving the system by moving the lattice sites back-and-forth in the lattice direction. Figure 1(b) illustrate the tight confinement limit. Figure 1(c) illustrates a 2D lattice in the weak confinement limit, where one has an array of cigar shaped traps. We consider square arrays, with the shaking oriented $45^{\rm o}$ from a lattice direction. These geometries are motivated by the experiments performed at Chicago \cite{ChinFloquet2013,ChinFloquet2014}. Using a kinetic model, we predict the scattering rate of bosons from a BEC as a function of the transverse confinement. Bilitewski and Cooper have performed a related study of the population dynamics in the Floquet realization of the Harper-Hofstader model \cite{Cooperarxiv2015}. In places where our studies overlap our results agree.\\

In section II, we introduce our model for analyzing the shaken lattice experiments. We also discuss the general formalism for obtaining the Floquet band structure. In section III, we use Fermi's golden rule to predict the scattering rate for bosons out of the BEC and obtain the stability phase diagram for a BEC loaded in a one-dimensional shaken lattice and in a shaken square lattice. Finally, we conclude with directions for future experiments.\\
 
\section{Model}
Energy is not conserved in a periodically driven system since the Hamiltonian, $H(t)$ is time dependent and work is being done on the system. However, one can map the problem onto a static one by viewing the system stroboscopically i.e. at times $0,T,2 T \ldots nT$ (cf. ref.\cite{polkovnikov2014floquet}). At these times, the evolution operator is given by :\\
\be
U(nT) = U(T)^{n} = {\mathcal{T}}\exp \left( -i \int_{0} ^{n T}\!dt \,\ H(t)/\hbar \right),
\ee
where ${\mathcal{T}}$ stands for the time-ordering operator. This structure allows us to define a time-independent effective Hamiltonian, $H_{\rm eff}$ via
\be
U(T) = e^{(-i H_{\rm eff}T/\hbar)} =   {\mathcal{T}}\exp \left( -i \int_{0} ^{T}\!dt \,\ H(t)/\hbar \right) 
\label{quasi}
\ee
This defines the effective Hamiltonian, $H_{\rm eff}$ and the eigenvalues of the effective Hamiltonian, the quasi-energies. It is evident from Eq.(\ref{quasi}) that the effective Hamiltonian is not unique. If $\epsilon$ is a quasi-energy, then $\epsilon + m \hbar \omega$ is also a quasi-energy where $m$ is an integer and $\omega = 2 \pi/T$. Clearly, the system has no unique thermodynamic ground state and the eventual occupations of various modes can only be determined by detailed modeling of the kinetics.\\

For our model, it suffices to use the rotating wave approximations (RWA) and we do not need the full Floquet formalism. While this simplifies the mathematics, the conceptual issues are unchanged.\\
\begin{figure}
\includegraphics[scale=0.465]{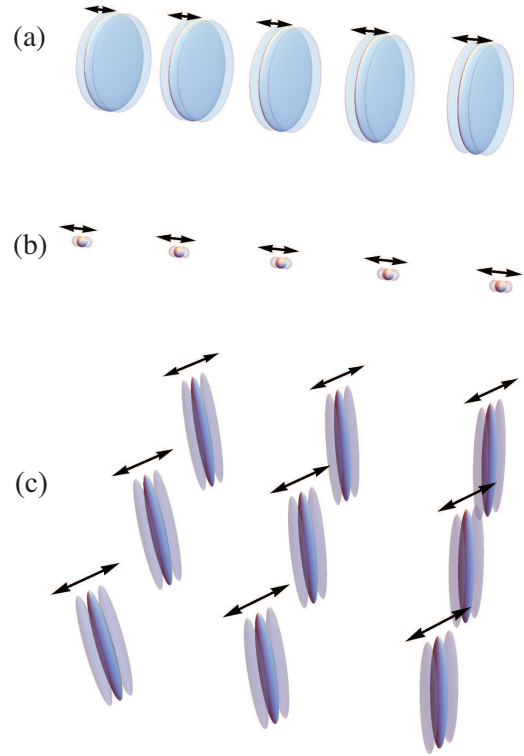}
\caption{(Color Online) Schematic of shaken optical lattices: (a)1D lattice with weak transverse confinement; (b) 1D Lattice  with tight transverse confinement; (c) 2D lattice with weak transverse confinement. Ellipsoids represent edges of cloud in each well of the optical lattice sites and arrows illustrate motion of trap. A typical spacing between lattice sites is 532 nm (half the laser wavelength $\lambda_L = 1064$ nm) and a typical shaking amplitude is 15 nm.}
\label{phase}
\end{figure} \\
\subsection{One-Dimensional Shaken Lattice}
\begin{figure}
\begin{center}
\includegraphics[scale=0.8]{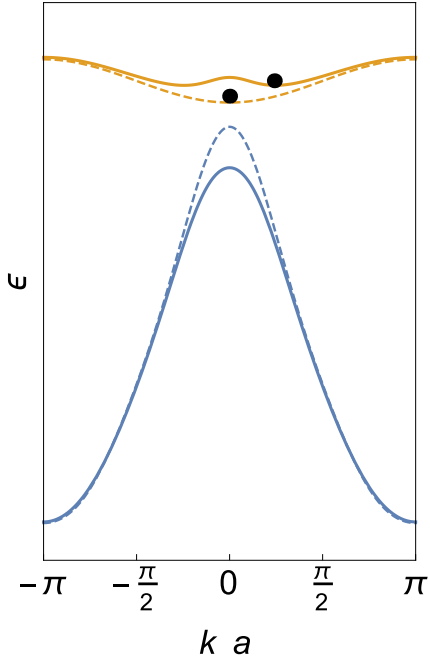}
\caption{(Color Online) Schematic showing first (top) and second (bottom) Floquet quasi-energy bands of an optical lattice: $\epsilon$ is the single-particle energy (arbitrary units used for schematic), $k$ is the quasi-momentum and $a$ is the lattice spacing. Since Floquet energies are only defined modulo the shaking quanta $\hbar \omega$,  the energy of the second band has been shifted down by $\hbar \omega$ so that it lies below the first band. Alternatively, this shift can be interpreted  as working in a dressed basis, where the energy includes a contribution from the phonons. The mixing between the bands depends on the shaking amplitude. Dashed curves correspond to weak shaking, where the first band has its minimum at $k = 0$. Solid curves correspond to strong shaking, where there are two minima at $k = \pm k_0 \ne 0$.}
\label{fig2}
\end{center}
\end{figure}

The starting point of our modeling is the set-up in \cite{ChinFloquet2013} where a BEC of $^{133}$Cs atoms is trapped in a one-dimensional shaken optical lattice (with weak transverse confinement). When the shaking amplitude exceeds a certain critical value, the BEC undergoes a phase transition to a ${\mathcal Z}_2$ superfluid (where condensation occurs at finite momentum $k = \pm k_0 \ne 0$). A schematic of the dispersion is shown in Fig. \ref{fig2}. For modeling this physics, it is sufficient to consider the first two Bloch bands and ignore the remaining bands (see supplement of \cite{ChinFloquet2013}).\\

In the frame of the moving lattice, the Hamiltonian for the driven system is given by $H=H_0(t) + H_{\rm int}$, where \cite{HolthausFloquetReview},
\bea
H_0(t) &=& \int\!d^3{\bf r}\,\ \Psi^{\dagger}({\bf r})\left(\frac{-\hbar^2}{2 m}\frac{d^2}{dz^2} + V_0 \sin^2\left(\frac{2 \pi z}{\lambda_L}\right) \right)\Psi({\bf r}) \nonumber\\
&+& \int\!d^3{\bf r} \,\ \Psi^{\dagger}({\bf r}) \left(z F_0 \cos(\omega t) \right)\Psi({\bf r}) \nonumber\\
&+& \int\!d^3{\bf r}\,\ \Psi^{\dagger}({\bf r})\left(\frac{-\hbar^2}{2 m}\nabla_{\myperp}^2 + m \Omega^2 (x^2+y^2) \right)\Psi({\bf r}), \nonumber \\
H_{\rm int}&=& \frac{g}{2} \int\!d^3{\bf r}\,\ \Psi^{\dagger} ({\bf r})  \Psi^{\dagger}({\bf r})  \Psi({\bf r})  \Psi({\bf r}).
\eea
The atomic mass is m, the wavelength of the laser forming the optical lattice is $\lambda_L$, the force from the periodic shaking is $F_0\cos(\omega t)$ and $g\approx \frac{4 \pi \hbar^2 a_s}{m}$ is the interaction strength, $a_s$ being the scattering length. The transverse trap frequency is $\Omega$.\\

As is detailed in Appendix A, the single particle part of the Hamiltonian describing the system $H_{\rm sp}$ can be written as
\bea
H_{\rm sp} &=&   \sum_{{\bf n},k} \epsilon^{(1)}_{{\bf n} k} a_{k}^{\bf n \dagger} a_{k}^{\bf n} + \epsilon^{(2)}_{{\bf n} k} b_{k}^{\bf n \dagger} b_{k}^{\bf n} \nonumber\\
&+& F_0 \cos(\omega t)\left(\chi a_{k}^{\bf n \dagger} b_{k}^{\bf n} +\chi ^{*} b_{k}^{\bf n \dagger} a_{k}^{\bf n} \right)
\eea
Here, $\epsilon^{(1)}_{{\bf n} k} (\epsilon^{(2)}_{{\bf n} k})$ is the dispersion of the first (second) band, $a_{k}^{\bf n} (b_{k}^{\bf n})$ is the annihilation operator for particles in the first (second) band with the harmonic oscillator level being ${\bf n}$ and $\chi$ is dipole matrix element between the first and the second band. As described in Appendix A, $\epsilon_{{\bf n}k}$ is generally time-dependent. However, when $F_0 a/(\hbar \omega) \ll 1$, $\epsilon_{{\bf n}k}$ can be taken to be time-independent. \\

We make the transformation $b_{\bf k} \rightarrow \exp(- i \omega t) b_{\bf k}$ and discard far off-resonant terms (making the rotating wave approximation) to simplify the single-particle Hamiltonian :
\bea
H_{\rm RWA} ^{(\rm sp)} &=& \sum_{{\bf n}, k} \epsilon^{(1)}_{{\bf n}k} a_{k} ^{{\bf n} \dagger} a_{k}^{\bf n} + \epsilon^{(2)}_{{\bf n}k}  b_{k} ^{{\bf n}\dagger} b_{k}^{\bf n} \nonumber \\
&+& \chi F\left(a_{k} ^{{\bf n}\dagger} b_{k}^{\bf n} +  b_{k} ^{{\bf n} \dagger} a_{k}^{\bf n}  \right),
\eea
Here $\epsilon^{(1)}_{{\bf n} k} = \epsilon^{(1)}_{k} + (n_x+n_y+1)\hbar \Omega$, $\epsilon^{(2)}_{{\bf n}k} = \epsilon^{(2)}_{k} + (n_x+n_y+1)\hbar \Omega - \hbar \omega$.  We diagonalize this quadratic form writing 
\be
H_{\rm RWA} ^{(\rm sp)} = \sum_{{\bf n} k} \overline{\epsilon}^{(1)}_{{\bf n} k} \overline{a}_{k} ^{{\bf n}\dagger} \overline{a}_{k}^{\bf n}+ \overline{\epsilon}^{(2)}_{{\bf n} k} \overline{b}_{k} ^{{\bf n}\dagger}  \overline{b}_{k}^{\bf n}
\label{hamrwa}
\ee
For a particular value of ${\bf n} = \{n_x,n_y\}$, the dressed dispersions $\overline{\epsilon}^{(1)}_{{\bf n} k}$ and $\overline{\epsilon}^{(2)}_{{\bf n} k}$ are shown as solid lines in Fig. \ref{fig2}. The bare dispersions $\epsilon^{(1)}_{{\bf n} k}$ and $\epsilon^{(2)}_{{\bf n} k}$ are shown as dashed lines. Reference \cite{ChoudhuryMuellerPRA2014} illustrates how the Floquet dispersion can be calculated when the RWA is not applicable.\\

\subsection{Shaken Square Lattice}
We can easily extend the analysis of the previous section to the case of the square lattice. Since the shaken square lattice is separable and equivalent to two shaken one-dimensional lattices, one can write down the single-particle part of the Hamiltonian, $H_{\rm 2D}$ in the frame of the optical lattice as :\\
\bea
H_{\rm 2D} &=& \int\!d^3{\bf r} \, \Psi({\bf r})^{\dagger}  (H_{\rm 1D} (z) + H_{\rm 1 D} (y)) \Psi({\bf r}) \nonumber \\
&+& \left (-\frac{\hbar^2}{2m} \frac{d^2}{dx^2} + m  \Omega^2 x^2\right) \Psi({\bf r})^{\dagger}\Psi({\bf r})
\eea
where,
\bea
H_{\rm 1D} (z) &=& -\frac{\hbar^2}{2m} \frac{d^2}{dz^2} + V_0 \sin^2\left(\frac{2 \pi z}{\lambda_L}\right) + z F_0 \cos(\omega t) \nonumber \\
H_{\rm 1D} (y)  &=&  -\frac{\hbar^2}{2m} \frac{d^2}{dy^2} + V_0 \sin^2\left(\frac{2 \pi y}{\lambda_L}\right) + y F_0 \cos(\omega t) \nonumber \\
\eea
Performing the same manipulations as in the last section, we end up with the following single particle Hamiltonian  :\\
\bea
H_{\rm 2D}^{\rm RWA} &=& \sum_{{n},{k_y,k_z}}\overline{\epsilon}^{(1,1)}_{{n}, k_x,k_y} \overline{a}_{k_x,k_y} ^{{ n}\dagger} \overline{a}_{k_x,k_y}^{n}+ \overline{\epsilon}^{(1,2)}_{{n}, k_x,k_y} \overline{b}_{k_x,k_y} ^{{n}\dagger}  \overline{b}_{k_x,k_y}^{n} \nonumber\\
&+& \overline{\epsilon}^{(2,1)}_{{n}, k_x,k_y} \overline{c}_{k_x,k_y} ^{{ n}\dagger} \overline{c}_{k_x,k_y}^{n}+ \overline{\epsilon}^{(2,2)}_{{n}, k_x,k_y} \overline{d}_{k_x,k_y} ^{{n}\dagger}  \overline{d}_{k_x,k_y}^{n} \nonumber\\
\label{hamrwa2D}
\eea
where $\overline{\epsilon}^{(i,j)}_{{n}, k_y,k_y} = \overline{\epsilon}^{(i)}_{{n=0}, k_z} + \overline{\epsilon}^{(j)}_{{n=0}, k_y} + (n+1/2) \hbar \Omega$. \\

\begin{figure}
\begin{center}
\includegraphics[scale=0.5]{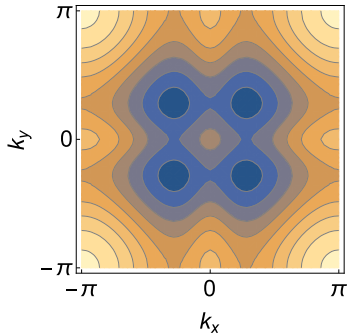}
\caption{(Color Online) Schematic showing the dispersion of the first Floquet band of a shaken square lattice beyond a critical amplitude. Color represents energy in units of the recoil energy, $E_R$ (see scale). We see that the superfluid order parameter develops a $D_4$ symmetry in momentum space.}
\label{fig2D}
\end{center}
\end{figure}

Due to the separability of the square lattice, instead of the $\mathcal{Z}_2$ reflection symmetry, the ground band develops a $D_4$ symmetry for shaking beyond a critical force. We show this schematically in Fig. \ref{fig2D}.\\

\section{Stability Analysis}
In this section, we use a kinetic approach to investigate the stability of a Floquet BEC as a function of transverse confinement. One would expect from our results in \cite{ChoudhuryMuellerPRA2014} that the Floquet BEC would be stable if the transverse confinement exceeds a critical value. For both the 1D shaken lattice and the shaken square lattice, we find this critical transverse confinement strength. For tighter potentials, the condensate is truly stable against energy-momentum conserving two-body collisions. We also identify several distinct signatures of interaction-driven scattering.   \\

Within our rotating wave approximation we can simply apply Fermi's golden rule. The rate of scattering of two atoms out of the BEC  is then given by: \\ 
\be
\frac{dN}{dt} = \frac{2 \pi}{\hbar} \sum_f\vert \langle \psi_f\vert H_{\rm int} \vert \psi_i \rangle \vert^2 \delta(\epsilon_f-\epsilon_i)
\label{scateqn}
\ee
where
\bea
\vert\psi_i\rangle &=& \frac{(\overline{a}_{k_0} ^{{\bf 0}\dagger})^N}{\sqrt{N !}} |0\rangle \nonumber \\
\vert\psi_f\rangle &=& \overline{\Psi}_{k_0+k} ^{\dagger} \overline{\Psi}_{k_0-k} ^{\dagger} \frac{(\overline{a}_{k_0} ^{{\bf 0}\dagger})^{(N-2)}}{\sqrt{N-2 !}} \vert 0\rangle \nonumber \\
\label{states}
\eea
where, $\overline{\Psi}_k$ is a shorthand for representing $\{\overline{a}_k,\overline{b}_k,\overline{c}_k,\overline{d}_k\}$,the state $\vert \psi_i \rangle$ denotes the BEC where the bosons have condensed at momentum $k_0$, while $\vert\psi_f\rangle$ denotes a state where two bosons have scattered out of the condensate to momenta $k_0+k$ and $k_0-k$ respectively. The energies of the final states are $\epsilon_f$ and $\epsilon_i$ respectively. If we did not use the Rotating Wave Approximation, a more complicated expression is necessary \cite{DemlerFloquetTransport}. Using Eq.(\ref{scateqn}) we investigate the stability of a Floquet BEC. All our calculations are done for the experimental parameters of Ref.\cite{ChinFloquet2013} a lattice depth of 7$E_R$, where the recoil energy, $E_R = h^2/(2m\lambda_L^2)$ where $\lambda_L = 1064$ nm and $m = 133$ amu. For these units, the zero-momentum bandgap for the 1D optical lattice is is 4.96 $E_R$ and the lattice is shaken at the blue detuned frequency of 5.5 $E_R$. It is reasonable to assume that loss is exponential. If not, Eq.(\ref{scateqn}) only describes the short-time behavior. At finite temperature, there are also heating processes involving one condensed atom and non-condensed atoms, or two non-condensed atoms. At typical BEC temperatures, these are negligible.
\subsection{One-Dimensional Shaken Lattice}
We first consider the case of a Floquet BEC loaded in a shaken 1D lattice. For this case, the boson scattering rate in Eq.(\ref{scateqn}) can be expressed as :
\be
\frac{dN}{dt} =\frac{2 \pi}{\hbar}\frac{g^2}{4} \frac{N^2}{L l_{\myperp}^{2}} \frac{1}{E_R a^3} \Gamma
\label{scatrate1d}
\ee
where $\Gamma$ is the adimensional scattering rate, $L$ is the linear system size and $l_{\myperp} = \sqrt{\hbar/(m \Omega)}$. The detailed derivation and the expression for $\Gamma$ are given in Appendix B. $\Gamma$ depends on the lattice depth, shaking frequency, shaking force and transverse confinement. It does not depend on the scattering length or the density.\\

\begin{figure}
\includegraphics[scale=0.48]{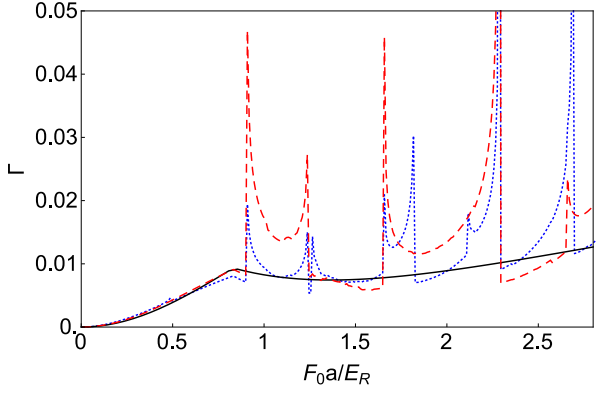}
\caption{(Color Online) Adimensional scattering rate $\Gamma$ as a function of the forcing amplitude, $F_0$ in the limit of weak confinement into a 1D lattice [Fig. 1(a)]. Blue, Dotted : $\hbar \Omega/E_R = 0.04$, Red, Dashed : $\hbar \Omega/E_R = 0.08$, Black, Solid: Analytic result from Ref.\cite{ChoudhuryMuellerPRA2015}.  }
\label{scattering}
\end{figure}

Fig. \ref{scattering} shows $\Gamma$ vs $F_0$ for weak transverse confinement ($\hbar \Omega/E_R = 0.04 \,{\rm and}\, 0.08$). For small $F_0$, $\Gamma$ rises quadratically and is roughly independent of $\Omega$. For large $F_0$, a series of resonances are visible.The lifetime of the condensate is given by :
\be
\tau = \frac{N}{dN/dt} = \frac{m L l_{\myperp}^2 a}{8 h a_s^2 N \Gamma}
\label{1Dlifetime}
\ee
Taking typical experimental parameters from the experiment in ref.\cite{ChinFloquet2013}, $m=133 \,{\rm amu}$, $L=30000 \, {\rm nm}$, $l_{\myperp}=1000 \, {\rm nm}$ $a_s = 1.5 \, {\rm nm}$, $N = 30,000$ and $\Gamma  = 0.01$, we get $\tau \sim 1 \, {\rm s}$. \\
\begin{figure}
\includegraphics[scale=0.43]{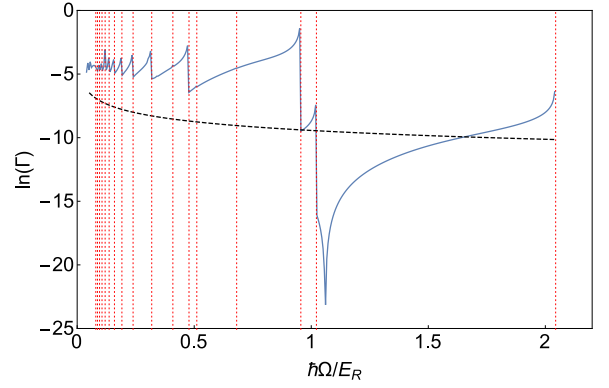}
\caption{(Color Online) Logarithm of the adimensional scattering rate, $\Gamma$ in a 1D lattice [Fig. 1(a),(b)] as a function of the transverse trapping frequency, $\Omega$ for a fixed value of the forcing amplitude, $F_0=F_c$, where $F_c$ is the amplitude where the dispersion of the ground band is quartic near $k=0$. Red vertical lines denote resonances at $\Omega = \Omega_n^{(a)}, \Omega_n^{(b)}$ corresponding to the closing of scattering channels (see text).The black dashed line shows the value of ln$(\Gamma)$ for different values of the transverse confinement for which the BEC lifetime is greater than 10 s (assuming the parameters quoted after Eq.(\ref{1Dlifetime})).}
\label{scat2}
\end{figure}
\begin{figure}
\includegraphics[scale=0.45]{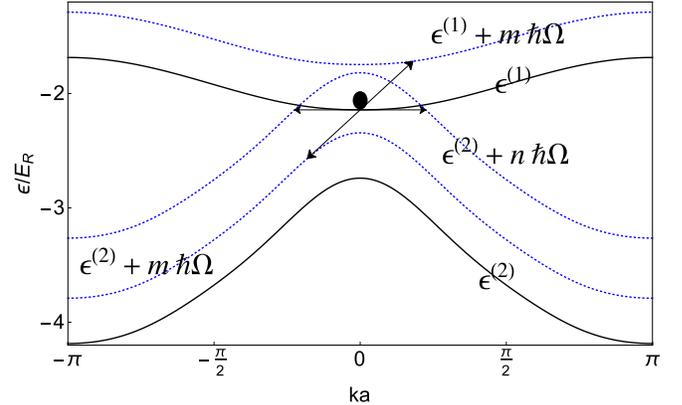}
\caption{(Color Online) Schematic illustrating conservation of energy and momentum in two-body collisions in a shaken pancake lattice. Black dot denotes condensate in first band at $k=0$. Solid lines show first and second with no transverse excitations. Arrows denote an energy and momentum conserving collision. The resonances in Fig. 4,5 correspond to the situation where the final states have $\vert ka \vert = \pi$}
\label{demo}
\end{figure}
\begin{figure}
\includegraphics[scale=0.45]{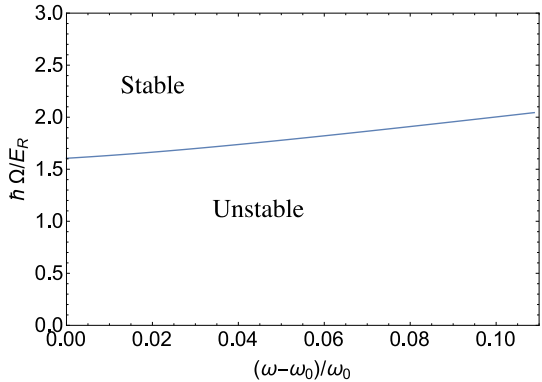}
\caption{(Color Online) Stability phase diagram for a BEC in a driven 1D lattice for a fixed value of the forcing amplitude, $F_0=F_c$. Here $\omega-\omega_0$ is the detuning of the shaking frequency $\omega$ from the zero-momentum bandgap $\omega_0$.}
\label{stability}
\end{figure}

To better understand the structure of resonances in Fig. \ref{scattering}, we plot $\Gamma$ on a log scale as a function of the transverse confinement frequency, $\Omega$ in Fig. \ref{scat2}. The set of vertical lines are given by the formula $\hbar \Omega_n^{(a)}=E_{12}/(2 n)$ and $\hbar \Omega_n^{(b)}=E_{22}/(2 n)$, where $E_{12} = \epsilon^{(1)}_{\pi}+\epsilon^{(2)}_{-\pi}-2 \epsilon^{(1)}_{0}$ and $E_{22} = \epsilon^{(2)}_{\pi}+\epsilon^{(2)}_{-\pi}-2 \epsilon^{(1)}_{0}$: these are related to the bandwidths of the two bands. These energy values, $E_{12}$ and $E_{22}$ correspond to the maximum longitudinal energy transfer in two different scattering channels and the resonance structure in Fig. \ref{scat2} corresponds to the closing of scattering channels. The factor of $2n$ corresponds to the spacing of parity allowed states. This structure can be understood by considering the energy and momentum conserving scattering processes in Fig.(\ref{demo}). The density of states is large when the final state has $\vert k a\vert = \pi$. This resonance structure leads to special parameters where the BEC would be particularly stable or unstable. These resonances are a useful fingerprint of the loss mechanism and can be used in an experiment to test our model of interaction-driven instability. The dashed line in Fig.\ref{scat2} corresponds to a lifetime of $\tau \approx 10$s (using the parameters below Eq.(\ref{1Dlifetime})). There is a large window around $\hbar\Omega \sim 1.1E_R$, where the lifetime exceeds 10 s. The BEC is completely stable against collisions $\hbar \Omega > 2.05 E_R$. In Fig. \ref{stability}, we show how the stability boundary varies with drive frequency. In terms of the dispersions of the two bands, the critical confinement is given by :\\
\be
\hbar \Omega = \epsilon^{(2)}_{\pi}+\epsilon^{(2)}_{-\pi}-2 \epsilon^{(1)}_{0}
\ee
For larger $\Omega$, energy and momentum can't be conserved in 2-body collisions.

\subsection{Two Dimensional Shaken Lattice}
In this section, we explore the stability of a Bose-Einstein condensate loaded in a two-dimensional optical lattice. The stability analysis is very similar to that of the shaken 1D lattice. The scattering rate of bosons can be written down as :
\be
\frac{dN}{dt} = \frac{2 \pi}{\hbar}\frac{g^2}{4} \frac{N^2}{L_y L_z l_{\myperp}} \frac{1}{E_R a^3} \Gamma
\label{scatrate2d}
\ee
where $\Gamma$ is the adimensional scattering rate and $L_z$ and $L_y$ denotes the linear system size in the z and y directions. The detailed derivation and the expression for $\Gamma$ are given in Appendix B. \\
\begin{figure}
\includegraphics[scale=0.55]{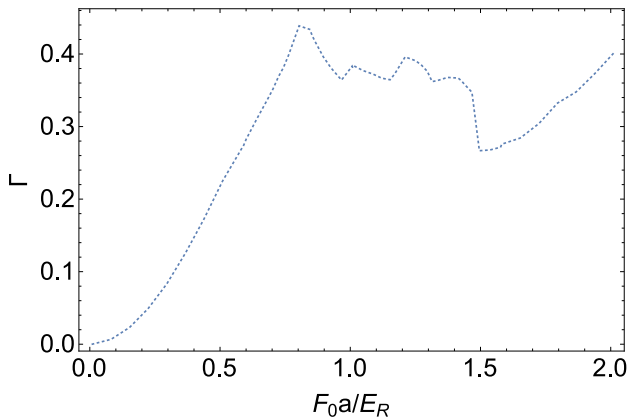}
\caption{(Color Online) Adimensional scattering rate $\Gamma$ as a function of the forcing amplitude, $F_0$ in the limit of weak confinement ($\hbar \Omega = 0.08 E_R$) for a 2D lattice [Fig. 1(c)]}
\label{scat2d}
\end{figure}
\begin{figure}
\includegraphics[scale=0.5]{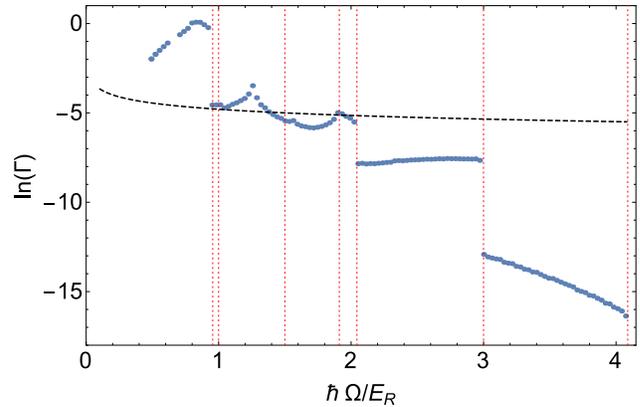}
\caption{(Color Online) Logarithm of the adimensional scattering rate $\Gamma$ in a 2D lattice as a function of the transverse confinement, $\Omega$ for a fixed value of the forcing amplitude, $F_0=F_c$. The black dashed line shows the value of ln$(\Gamma)$ for different values of the transverse confinement for which the BEC lifetime is greater than 10 s (assuming the parameters quoted after Eq.(\ref{2Dlifetime})).}
\label{scat2_2}
\end{figure}\\

We show $\Gamma$ for a relatively weak value of transverse confinement ($\hbar \Omega/E_R = 0.08$) in Fig. \ref{scat2d}.  We see that the adimensional scattering rate, $\Gamma$, is higher for the shaken two-dimensional square lattice when compared to the one-dimensional lattice. \\

The lifetime of the condensate is given by :
\be
\tau = \frac{N}{dN/dt} = \frac{m L_y L_z l_{\myperp} a}{8 h a_s^2 N \Gamma}
\label{2Dlifetime}
\ee
Now, taking typical experimental parameters from the experiment in ref.\cite{ChinFloquet2013}, $m=133 \,{\rm amu}$, $L_y=30000 \, {\rm nm}$, $L_z=30000 \, {\rm nm}$, $l_{\myperp}=1000 \, {\rm nm}$ $a_s = 1.5 \, {\rm nm}$, $N = 30,000$ and $\Gamma  = 0.4$, we get $\tau \sim 0.73 \, {\rm s}$. \\

The scatter of points in Fig. {\ref{scat2d} is related to the resonances. Again, these can be explored by fixing $F_0$ to some value (here, $F_c$) and then plotting the scattering rate, $\Gamma$ as a function of the transverse confinement, $\Omega$ as shown in Fig. {\ref{scat2_2}. The black dashed line again corresponds to a lifetime of 10 s. There are specific value of $\Omega$ at which the scattering rate drops significantly. These values of $\Omega$ are shown as vertical lines in Fig. {\ref{scat2_2} and correspond to $\hbar \Omega = (E_{12}+E_{22})/(2 n), (E_{12}+E_{12})/(2 n),(E_{22}+E_{22})/(2 n)$. As in the 1D case, these frequencies correspond to the closing of scattering channels. There is also structure related to the van Hove singularities in the density of states, but for clarity, we do not mark them with vertical lines. Beyond a transverse confinement of 1.4 $E_R$, the BEC will almost always have a lifetime $\tau > 10$s.\\

Due to the separability of the Hamiltonian, the critical transverse confinement for the 2D square lattice is exactly twice that of the one-dimensional lattice, so the stability phase diagram is readily inferred from Fig.\ref{stability}\\

\section{Conclusion}
In this paper we studied the effect of transverse confinement on the stability of a Floquet BEC for both a shaken 1D lattice and a shaken 2D square lattice. We obtained scattering rates as well as the stability phase diagrams for both systems. The scattering rate shows a resonant structure and fine tuning parameters can drastically reduce the loss rate. This structure arises from the opening and closing of loss channels corresponding to the quantized transverse modes. It provides a fingerprint of the loss mechanism and could be a valuable tool for minimizing loss. We find a critical value of transverse confinement, beyond which there are no allowed 2-body scattering processes which can deplete the condensate. Well before this point however, the scattering rate drops to extremely small values, making the BEC stable for the time-scales of the experiment.\\

The loss mechanism that we study has another distinct signature - namely that energy is converted from the time-dependent potential into transverse motion of the atoms. This transverse motion can be directly probed in time-of-flight experiments.

\section*{Acknowledgements}
We thank the Chin group (Cheng Chin, Harry Ha, Logan Clark and Colin Parker) for correspondence and discussions about their experiments. We especially thank Colin Parker for helping us find errors in our numerics. We acknowledge support from ARO-MURI Non-equilibrium Many-body Dynamics grant (W911NF-14-1-0003).\\

\appendix
\section{Derivation of the 1D Hamiltonian}
In a tight-binding prescription, the single-particle Hamiltonian describing the system in the frame co-moving with the lattice can be written as $H_0 (t)$:
\bea
H_0(t) &=& \int d^2 r_{\myperp} \sum_{ij} \left(-t_{ij} ^{(1)} a_{i}^{\dagger}a_{j} +  t_{ij} ^{(2)}  b_{i}^{\dagger}b_{j} + h.c.\right) \nonumber \\
&+&  \sum_{j} F(t)  \left(z_j \left(a_{j}^{\dagger} a_{j} + b_{j}^{\dagger} b_{j}\right) + \chi_j a_j^{\dagger} b_j + \chi_j ^{*} b_j^{\dagger} a_j \right) \nonumber\\
&+& \frac{\hbar^2}{2m} \left(\nabla_{\myperp} a_j^{\dagger} \nabla_{\myperp} a_j + \nabla_{\myperp} b_j^{\dagger} \nabla_{\myperp} b_j \right) \nonumber\\
&+& m \Omega^2 (x^2+y^2)\left (a_j^{\dagger}a_j +b_j^{\dagger}b_j\right)\\
\eea
where,
\bea
\chi_j  &=& \int\!dz \,\ z w_1^{*}(z-z_j) w_2(z-z_j) \nonumber \\
t_{ij} ^{(1)} &=& \int dz\,\ w_1^{*}(z-z_i)\left(\frac{-\hbar^2}{2 m} \frac{d^2}{dz^2}+ V(z)\right)w_1^{*}(z-z_j)\nonumber \\
t_{ij} ^{(2)} &=& \int dz\,\ w_2^{*}(z-z_i) \left(\frac{-\hbar^2}{2 m} \frac{d^2}{dx^2}+V(z)\right)w_2^{*}(z-z_j) \nonumber \\
F(t) &=& F_0\cos(\omega t) 
\eea
Here, $w_i$ is the Wannier function for the $i$th band. It should be noted that $\chi_j$ is independent of $j$ and so we can call it $\chi$. The operators $a_j$ and $b_j$ annihilate particles in the two bands. If necessary more bands can be included. \\

Performing a basis rotation : $\vert\psi\rangle \rightarrow U_{c}(t) \vert \psi\rangle$ where
\be
U_{c}(t) = \exp\left(- \frac{i}{\hbar} \int_{0}^{t} \sum_{j}z_ j F_0\cos(\omega t) (a_{j}^{\dagger} a_{j} + b_{j}^{\dagger} b_{j}) \right),
\label{unitary}
\ee
we transform the Hamiltonian as:
\bea
H_0^{\prime} (t) &=& U_{c}H_0(t)U_{c}^{-1} - i \hbar U_{c}\partial_t U_{c}^{-1} \nonumber  \\
&=& \sum_{n_x,n_y} H_{\bf n} 
\eea
with
\bea
H_{\bf n}&=& \sum_{ij} \left(-J_{ij} ^{(1)} (t) a_{i}^{\bf n \dagger}a_{j}^{\bf n} +  J_{ij} ^{(2)}(t)  b_{i}^{\bf n \dagger}b_{j}^{\bf n} + h.c.\right) \nonumber\\
&+&\sum_j F \cos(\omega t)  \left(\chi a_{j}^{\bf n \dagger} b_{j}^{\bf n} +\chi ^{*} b_{j}^{\bf n \dagger} a_{j}^{\bf n} \right) \nonumber\\
&+& \sum_{\bf n} \hbar \Omega (n_x+n_y+1) \left( a_j^{\bf n \dagger} a_j^{\bf n} +  b_j^{\bf n \dagger} b_j^{\bf n} \right)\nonumber \\
&=& \sum_k \sum_m  \cos(m k a)\left(-J_{m} ^{(1)} (t) a_{k}^{\bf n \dagger}a_{k}^{\bf n} -J_{m} ^{(2)} (t) b_{k}^{\bf n \dagger}b_{k}^{\bf n}\right)\nonumber \\
&+& \sum_k F_0\cos(\omega t)  \left(\chi a_{k}^{\bf n \dagger} b_{k}^{\bf n} +\chi ^{*} b_{k}^{\bf n \dagger} a_{k}^{\bf n} \right) \nonumber\\
&+& \sum_{\bf  n}  \hbar \Omega (n_x+n_y+1) \left( a_k^{\bf n \dagger} a_k^{\bf n} +  b_k^{\bf n \dagger} b_k^{\bf n} \right)\nonumber \\
\eea
where,
\bea
J_{ij}^{\sigma} (t) &=& t_{ij}^{\sigma} \exp(-i F_0 \frac{\sin(\omega t)}{\hbar \omega} (z_i-z_j)) \nonumber\\
&=& t_{ij}^{\sigma} \exp(-i F_0 \frac{\sin(\omega t)}{\hbar \omega} a (i-j)) ,
\label{rwa1}
\eea
$a = \lambda_L/2$ is the lattice spacing and $\chi=\chi^{*}$ for a suitable choice of phase for $a_k$ and $b_k$. We use ${\bf n}$ as a shorthand for denoting $\{n_x,n_y\}$.\\

In the limit of $F_0 a/(\hbar \omega) \ll 1$, $J_{ij}^{\sigma} (t) = t_{ij}^{\sigma}$. Hence, we can write down the Hamiltonian as :
\bea
H_{\rm sp} &=&   \sum_{{\bf n},k} \epsilon^{(1)}_{{\bf n} k} a_{k}^{\bf n \dagger} a_{k}^{\bf n} + \epsilon^{(2)}_{{\bf n} k} b_{k}^{\bf n \dagger} b_{k}^{\bf n} \nonumber\\
&+& F_0 \cos(\omega t)\left(\chi a_{k}^{\bf n \dagger} b_{k}^{\bf n} +\chi ^{*} b_{k}^{\bf n \dagger} a_{k}^{\bf n} \right) 
\eea
where,
\bea
\epsilon^{(1)}_{{\bf n} k} &=& \sum_k \sum_m  -t_{m} ^{(1)} \cos(m k a)+\hbar \Omega (n_x+n_y+1)\nonumber\\
\epsilon^{(2)}_{{\bf n} k} &=& \sum_k \sum_m  t_{m} ^{(2)} \cos(m k a)+\hbar \Omega (n_x+n_y+1) \nonumber\\
\eea

\section{Derivation of the scattering rate}
\subsection{1D Lattice}

For the case of the 1D optical lattice, the scattering rate in Eq.(\ref{scateqn}) can be written down as 

\be
\frac{dN}{dt} = \frac{2 \pi}{\hbar}\frac{g^2}{4} N^2 \frac{L}{2 \pi} \sum_{{\bf n}_a,{\bf n}_b} \int\!dk \Gamma^{{\bf n}_a, {\bf n}_b}_k \delta(\epsilon_f-\epsilon_i)
\label{rate1}
\ee
where
\be
\Gamma^{{\bf n}_a,{\bf n}_b}_k =\left|\frac{ I_x^{{\bf n}_a,{\bf n}_b}I_y^{{\bf n}_a,{\bf n}_b}\langle \psi_f \vert\int\! dk \Psi_{k_0-k}^{\dagger}\Psi_{k_0+k}^{\dagger}\Psi_{k_0}\Psi_{k_0}\vert \psi_i \rangle}{L l_{\myperp}^2}\right| ^2
\nonumber
\ee
and
\be
 I_x^{{\bf n}_a,{\bf n}_b} = \int dx \phi^{(n_a^x)}(x)\phi^{(n_b^x)}(x)\phi^{(0)}(x)\phi^{(0)}(x)
\label{Ieqn}
\ee
with ${\bf n}_a({\bf n}_b) = \{n_a^x,n_a^y\}( \{n_b^x,n_b^y\})$, $\phi^{(n)} (x) = H_n(x) \exp(-x^2/2)$, $H_n(x)$ being the Hermite polynomial of order n. An important consequence of the form of \ref{Ieqn} is that $I_x^{{\bf n}_a,{\bf n}_b}=0$ unless $n_a^x$ and $n_b^x$ (as well as $n_a^y$ and $n_b^y$) have the same parity. Finally, Eq.(\ref{rate1}) can be simplified to write

\be
\frac{dN}{dt} =\frac{2 \pi}{\hbar}\frac{g^2}{4} \frac{N^2}{L l_{\myperp}^{2}} \frac{1}{E_R a^3} \Gamma
\ee \\
with 
\be
\Gamma = \frac{L^2 l_{\myperp}^2 E_R a^3}{2 \pi}\sum_{{\bf n}_a,{\bf n}_b} \int\!dk \Gamma^{{\bf n}_a, {\bf n}_b}_k \delta(\epsilon_f-\epsilon_i) \nonumber
\ee
This is Eq.(\ref{scatrate1d}) in the main text. \\
\subsection{2D Square Lattice}
For the case of the 2D square lattice, the scattering rate in Eq.(\ref{scateqn}) can be written down as 
\be
\frac{dN}{dt} =\frac{2 \pi}{\hbar} \frac{g^2}{4} {N^2}\frac{L_yL_z}{(2 \pi)^2} \sum_{n_a,n_b} \int d^2{\bf k} \Gamma^{n_a,n_b}_{\bf k} \delta(\epsilon_f-\epsilon_i)
\label{rate2d}
\ee
where
\be
\Gamma^{{n}_a,{n}_b}_{\bf k} =\left\vert \frac{ I^{{n}_a,{n}_b}\langle \psi_f \vert\int\! d^2{\bf k} \Psi_{\bf k_0-k}^{\dagger}\Psi_{\bf k_0+k}^{\dagger}\Psi_{\bf k_0}\Psi_{\bf k_0}\vert \psi_i \rangle}{L_yL_z l_{\myperp}} \right\vert^2
\nonumber
\ee
and
\be
I^{n_a,n_b} = \int dx \phi^{(n_a)}(x)\phi^{(n_b)}(x)\phi^{(0)}(x)\phi^{(0)}(x)
\label{Ieqn2}
\ee
with $l_{\myperp} = \sqrt{\hbar/(m \Omega)}$ just as in the case of the 1D shaken lattice. An important consequence of the form of \ref{Ieqn2} is that $I^{n_a,n_b}$ is 0 unless $n_a$ and $n_b$ have the same priority. Eqn.(\ref{rate2d}) simplifies to give :
\be
\frac{dN}{dt} = \frac{2 \pi}{\hbar}\frac{g^2}{4} \frac{N^2}{L_y L_z l_{\myperp}} \frac{1}{E_R a^3} \Gamma
\ee
with
\be
\Gamma = \frac{L_y^2 L_z^2 l_{\myperp} E_R a^3}{(2 \pi)^2} \sum_{n_a,n_b} \int d^2{\bf k} \Gamma^{n_a,n_b}_{\bf k} \delta(\epsilon_f-\epsilon_i)
\nonumber
\ee
This is Eq.(\ref{scatrate2d}) in the main text.

\end{document}